\documentclass[a4paper,12pt]{article}
\usepackage{float}
\usepackage[margin=2.5cm]{geometry}
\usepackage{setspace}
\usepackage{indentfirst}
\usepackage{footmisc}
\usepackage{amsmath,amssymb,mathtools}
\usepackage{graphicx}
\usepackage{booktabs}
\usepackage{tikz}
\usetikzlibrary{arrows.meta,calc}
\usepackage[nosort]{cite}
\usepackage[svgnames]{xcolor}
\usepackage[colorlinks=true,allcolors=.,bookmarksnumbered=true,pdfpagemode=UseNone,pdfstartview=FitH]{hyperref}
\usepackage{microtype}

\newcommand{\dif}{\mathrm{d}}
\newcommand{\Ast}{\mathrm{Ast}}
\newcommand{\OP}{\mathrm{OP}}
\newcommand{\BR}{\mathrm{BR}}

\numberwithin{equation}{section}
\makeatletter
\renewcommand\section{\@startsection {section}{1}{\z@}{-3.5ex \@plus -1ex \@minus -.2ex}{2.3ex \@plus.2ex}{\normalfont\large\sffamily\bfseries}}
\renewcommand\subsection{\@startsection{subsection}{2}{\z@}{-3.25ex\@plus -1ex \@minus -.2ex}{1.5ex \@plus .2ex}{\normalfont\normalsize\sffamily\bfseries}}
\makeatother

\begin{document}
\thispagestyle{empty}
\vbox{}\vspace{0.5cm}
\begin{center}
{\sffamily\LARGE The subcritical accelerating Schwarzschild--Bertotti--Robinson family}\\[10mm]
{\sffamily\large Kenneth Hong}\\[4mm]
{\sffamily\slshape Department of Physics, National University of Singapore, Singapore}\\[8mm]
\end{center}
\vspace{0.7cm}
\centerline{\sffamily\bfseries Abstract}
\bigskip
\noindent
We identify the general finite-$\ell$ neutral Alekseev--Astorino Schwarzschild--Bertotti--Robinson family with the subcritical branch of the independently obtained accelerating Schwarzschild--Bertotti--Robinson solution.  The exact scale-free dictionary is
\begin{equation*}
 B^2M^2=\frac{m^2}{b^2+\ell^2},\qquad
 AM=\frac{m\ell}{b^2+\ell^2},\qquad
 \frac{A}{B}=\frac{\ell}{\sqrt{b^2+\ell^2}},
\end{equation*}
up to the north--south orientation convention.  Thus finite real $\ell$ covers precisely the acceleration interval $|A|<|B|$: for $\ell\ne0$ the corresponding OP acceleration parameter is nonzero and the two axial openings are unequal, although no acceleration horizon is present.  We establish the identification by canonical Weyl-rod matching and horizon-shape comparison, an independent conicity check, and an exact off-horizon reconstruction of the metric and Maxwell field.  The centered limit $\ell=0$ is the nonaccelerating Schwarzschild--BR solution, whereas $|\ell|\to\infty$ approaches the critical boundary $|A/B|=1$ from below.  The Astorino parameter $\ell$ therefore parametrizes subcritical acceleration rather than an additional independent neutral black-hole degree of freedom.

\newpage
\section{Introduction}
The Bertotti--Robinson (BR) spacetime is a homogeneous Einstein--Maxwell solution locally isometric to $\mathrm{AdS}_2\times S^2$~\cite{Bertotti1959,Robinson1959}.  Its electromagnetic field and non-asymptotically flat geometry provide a setting for black holes distinct from the usual asymptotically flat and Melvin backgrounds.  Alekseev and Garc\'ia constructed a Schwarzschild black hole in this background~\cite{AlekseevGarcia1996}.  The solution and its limiting geometries were subsequently investigated in different forms, including the ultrarelativistic construction of Ortaggio and Astorino~\cite{OrtaggioAstorino2018}.

A related line of work concerns nonaligned Einstein--Maxwell fields of Petrov type D.  Van den Bergh and Carminati studied the nonaligned Robinson--Trautman sector~\cite{VanDenBerghCarminati2020}.  Ovcharenko and Podolsk\'y (OP) subsequently constructed a broader rotating family with nonaligned electromagnetic field~\cite{OP2025}; their analysis of its nontwisting sector identifies, after reparametrization, both the neutral Schwarzschild/C-metric family in BR and a distinct charged Reissner--Nordstr\"om--BR family~\cite{OP2026}.  They discuss connections with the Alekseev--Garc\'ia and other earlier solutions, although such relationships are not all given as explicit transformations between the respective coordinates and parameters.  Their later uniqueness result applies within a specified conformal-to-Carter ansatz and should not be interpreted as an unrestricted classification of all Einstein--Maxwell spacetimes~\cite{OPUniqueness2026}.

Further BR black-hole solutions have been obtained through Harrison transformations and related solution-generating methods.  Astorino obtained solutions in combined Bertotti--Robinson--Bonnor--Melvin fields~\cite{Astorino2025} and subsequently studied a Reissner--Nordstr\"om family in the same setting~\cite{Astorino2026}.  Di Pinto introduced a further parametrization of a nonaccelerating Schwarzschild--BR sector of the OP solutions and its Harrison-generated extension~\cite{DiPinto2026}.  Alekseev considered both a distinct ``geodesic''-motion construction~\cite{Alekseev2017} and charged acceleration in a homogeneous BR electric field~\cite{Alekseev2025}; Harrison and inversion transformations have also been applied to obtain vacuum descendants of BR configurations~\cite{Barrientos2026}.  These results concern related backgrounds and solution families, but the existence of common limits does not alone establish equivalence of the complete Einstein--Maxwell fields.

In Astorino's representation the neutral solution involves two spheroidal coordinate systems on the Weyl half-plane, with reference locations $\beta_1$ and $\beta_2$.  Their difference
\begin{equation}
 \ell\equiv\beta_2-\beta_1
 \label{eq:ell-beta-intro}
\end{equation}
is unchanged by a common translation of the Weyl coordinate.  Astorino analyzes the geometrical and axial-regularity consequences of the relative placement and also exhibits a coordinate equivalence between Schwarzschild--BR representations in the centered specialization $\beta_1=\beta_2=0$~\cite{Astorino2026}.  That transformation does not establish the relation of the general neutral finite-$\ell$ sector to the OP acceleration parameter.

The distinction matters because the finite-$\ell$ neutral solution has unequal axial conicities when $m\ell\ne0$, yet its Weyl representation has no separate acceleration-horizon rod.  In the OP representation the neutral radial function contains the factor $1+(B^2-A^2)R^2$: the subcritical interval $|A|<|B|$ has no additional real root, whereas additional roots occur for $|A|>|B|$.  The possibility of acceleration without a separate acceleration horizon is familiar in curved backgrounds, including the AdS C-metric~\cite{DiasLemos2003}.  The question here is whether the finite-$\ell$ Astorino geometry is precisely the subcritical OP geometry.  The identification requires a comparison of the complete Einstein--Maxwell fields.

We establish the local equivalence of the complete finite-real-$\ell$ neutral Alekseev--Astorino family with the subcritical accelerating Schwarzschild--BR family of OP.  The scale-free parameter correspondence is
\begin{equation}
 B^2M^2=\frac{m^2}{b^2+\ell^2},\qquad
 AM=\frac{m\ell}{b^2+\ell^2},
 \label{eq:intro-dict}
\end{equation}
and hence, with the positive-$B$ convention,
\begin{equation}
 \frac{A}{B}=\frac{\ell}{\sqrt{b^2+\ell^2}}.
 \label{eq:intro-ratio}
\end{equation}
Thus finite real $\ell$ covers precisely $-1<A/B<1$, with the centered member corresponding to $A=0$ and the critical boundary reached only as $|\ell|/b\to\infty$.  The parameter $\ell$ therefore parametrizes the subcritical acceleration degree of freedom, rather than an additional independent parameter of the neutral black hole.

The proof proceeds by matching the canonical Weyl black-hole rods, deriving the candidate dictionary from the intrinsic horizon shape, and checking it against the axial conicities.  We then reconstruct the second Astorino spheroidal system and give an explicit off-horizon transformation of the metric and magnetic Maxwell field.  The identification is local, subject to constant Killing-coordinate normalizations and electromagnetic conventions, and does not identify arbitrary global conical quotients.  The charged sector, the critical and supercritical regimes, and global causal extensions are not established here.

\section{The Alekseev--Astorino and Ovcharenko--Podolsk\'y solutions}\label{sec:two}
\subsection{The Alekseev--Astorino solution}
We consider the neutral sector of the Alekseev--Astorino solution of Ref.~\cite{Astorino2026}, obtained by setting
\begin{equation}
 \nu=0,\qquad \mu=\sigma=m,\qquad \hat b=b,
 \label{eq:neutral-specialization}
\end{equation}
while leaving the two Weyl reference locations $\beta_1$ and $\beta_2$ unrestricted.

Two spheroidal coordinate systems are used on the canonical Weyl half-plane:
\begin{align}
 z_A-\beta_1&=x_1y_1,&
 \rho_A^2&=(b^2+x_1^2)(1-y_1^2),
 \label{eq:br-spheroid-general}\\
 z_A-\beta_2&=x_2y_2,&
 \rho_A^2&=(x_2^2-m^2)(1-y_2^2).
 \label{eq:rod-spheroid-general}
\end{align}
Their translation-invariant separation will be denoted
\begin{equation}
 \ell\equiv\beta_2-\beta_1.
 \label{eq:ell-beta}
\end{equation}
In Weyl coordinates $(t_A,\rho_A,z_A,\phi_A)$, the metric can be written, with a slight change of notation from Ref.~\cite{Astorino2026}, as
\begin{equation}
 \dif s_A^2=f_A\,\dif t_A^2+
 \frac{g_A b^2}{\Delta_A}\left(\dif\rho_A^2+\dif z_A^2\right)
 -\frac{\rho_A^2}{f_A}\,\dif\phi_A^2,
\label{eq:Ast-neutral-metric}
\end{equation}
where
\begin{align}
 f_A&=-\frac{(b^2+x_1^2)(x_2+my_1)^2}
 {b^2(x_2^2-m^2)},
 \label{eq:Ast-neutral-f}\\
 g_A&=\frac{f_0D_0^2}{x_2^2-m^2y_2^2},
 \label{eq:Ast-neutral-g}\\
 \Delta_A&=\sqrt{\left[b^2+\rho_A^2+(z_A-\beta_1)^2\right]^2-4b^2\rho_A^2},
 \label{eq:Ast-neutral-Delta}\\
 D_0&=(\ell^2+b^2+m^2)(x_2+my_1)
 -2m\left[\ell(x_1-my_2)+(b^2+m^2)y_1\right].
 \label{eq:Ast-neutral-D}
\end{align}
We use the signature $(-+++)$, for which $f_A<0$ in the static exterior. The constant $f_0$ fixes the Weyl conformal normalization and is determined once the axial normalization and period of $\phi_A$ are specified.

The electric potential of the neutral Alekseev--Astorino solution can be written as
\begin{equation}
 A^{\rm Ast}_{\rm el}=-\frac{x_1+my_2}{b}\,\dif t_A.
\label{eq:Ast-neutral-electric}
\end{equation}
For comparison with the magnetic Ovcharenko--Podolsk\'y solution of Ref.~\cite{OP2026}, we perform an electromagnetic duality transformation,
\begin{equation}
 F^{\rm Ast}_{\rm mag}={}^\star F^{\rm Ast}_{\rm el},
 \qquad F^{\rm Ast}_{\rm el}=\dif A^{\rm Ast}_{\rm el},
\end{equation}
which leaves the metric unchanged. A potential for the dual magnetic field is
\begin{equation}
 A^{\rm Ast}_{\rm mag}=A^{\rm Ast}_\phi\,\dif\phi_A,
 \qquad
 A^{\rm Ast}_\phi=b\,\frac{(1-y_1)(x_2-m)}{x_2+my_1}+A_{\phi0}^{\rm Ast},
\label{eq:Ast-neutral-magnetic}
\end{equation}
where $A_{\phi0}^{\rm Ast}$ is an additive gauge constant. The sign of the dual field depends on the spacetime orientation.

\begin{figure}[t]
\centering
\begin{tikzpicture}[x=1cm,y=1cm,>=Latex,font=\small]
\draw[->,thin] (-4.6,0)--(4.8,0) node[right,font=\scriptsize] {$z_A$};
\coordinate (b1) at (-2.8,0); \coordinate (b2) at (0.9,0);
\fill (b1) circle (2.2pt); \fill (b2) circle (2.2pt);
\node[below=5pt,font=\scriptsize,align=center,text width=2.6cm] at (-3.15,0) {$\beta_1$\\BR-adapted Weyl reference};
\node[below=5pt,font=\scriptsize,align=center,text width=2.4cm] at (1.35,0) {$\beta_2$\\black-hole rod center};
\draw[very thick] (-0.15,0)--(1.95,0);
\fill (-0.15,0) circle (1.9pt); \fill (1.95,0) circle (1.9pt);
\node[above=5pt,font=\scriptsize] at (-0.15,0) {$\beta_2-m$};
\node[above=5pt,font=\scriptsize] at (1.95,0) {$\beta_2+m$};
\node[above=27pt,font=\scriptsize] at (0.9,0) {black-hole horizon};
\node[above=17pt,font=\scriptsize] at (-1.25,0) {$\mathcal A_-$};
\node[above=17pt,font=\scriptsize] at (3.15,0) {$\mathcal A_+$};
\draw[<->,thick] (-2.8,-1.65)--(0.9,-1.65);
\node[below,font=\scriptsize] at (-0.95,-1.65) {$\ell=\beta_2-\beta_1$};
\node[font=\scriptsize,align=center] at (0,-2.55) {a common Weyl translation changes $\beta_1$ and $\beta_2$\\but leaves their separation $\ell$ unchanged};
\end{tikzpicture}
\caption{Weyl structure of the neutral finite-$\ell$ Alekseev--Astorino solution before fixing the arbitrary axial origin.  The black-hole horizon is the single finite horizon rod centered at $\beta_2$.  The point $\beta_1$ labels the reference of the BR-adapted oblate system; it is not assumed to be an invariant ``center'' of the homogeneous BR universe.  The translation-invariant datum is $\ell=\beta_2-\beta_1$.  The two exterior axis segments are $\mathcal A_-$ and $\mathcal A_+$.}
\label{fig:astorino-rod}
\end{figure}
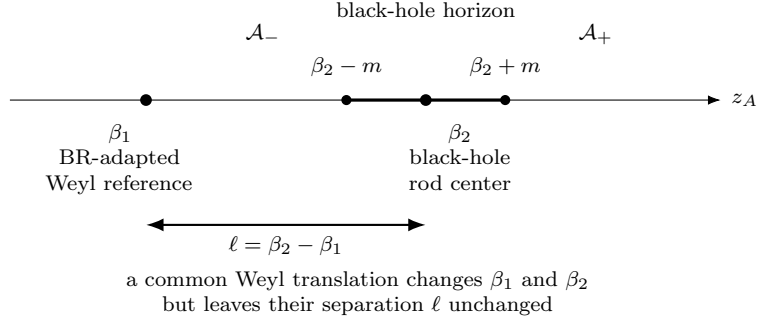

In the prolate coordinates, $x_2=m$ gives $\rho_A=0$ and $z_A-\beta_2=my_2$.  The corresponding finite black-hole horizon rod is
\begin{equation}
 \mathcal H_b^{\rm Ast}:\qquad \beta_2-m\le z_A\le\beta_2+m.
 \label{eq:ast-rod-general}
\end{equation}
Thus the neutral solution contains a single finite horizon rod, centered at $\beta_2$, while the oblate system is referenced to the distinct point $\beta_1$.  The corresponding Weyl structure is shown in Fig.~\ref{fig:astorino-rod}.

In Astorino's parametrization of the parent solution, $\mu$, $\nu$ and $\hat b$ are associated with mass, intrinsic electromagnetic charge and the external BR field, respectively.  The parameters $\beta_1$ and $\beta_2$ specify the two Weyl reference locations, whose separation is $\ell=\beta_2-\beta_1$.  Astorino interprets this separation geometrically in terms of the relative placement of the black hole and the BR background~\cite{Astorino2026}.  The explicit comparison with Schwarzschild--BR in spherical coordinates is restricted to $\beta_1=\beta_2=0$, and hence $\ell=0$.

The BR background is homogeneous and has no preferred spatial center.  Thus $\ell$ should not be interpreted directly as a proper distance from a distinguished center of the BR universe.  Under a common Weyl translation,
\begin{equation}
 z_A\mapsto z_A+z_0,\qquad
 \beta_1\mapsto\beta_1+z_0,\qquad
 \beta_2\mapsto\beta_2+z_0,
\end{equation}
the separation $\ell$ remains unchanged.  It therefore characterizes the relative placement of the two coordinate systems, although its physical interpretation is not yet apparent.

The finite black-hole rod separates the Weyl axis $\rho_A=0$ into two exterior segments, $\mathcal A_-:z_A<\beta_2-m$ and $\mathcal A_+:z_A>\beta_2+m$, shown in Fig.~\ref{fig:astorino-rod}.  Near either axis, the transverse metric takes the form
\begin{equation}
 \dif s_\perp^2\simeq \dif r_\perp^2+
 \kappa_\pm^2 r_\perp^2\,\dif\phi_A^2,
\end{equation}
where $\kappa_\pm$ are the corresponding conicity factors.  Defining
\begin{equation}
 K_+=b^2+(\ell+m)^2,\qquad
 K_-=b^2+(\ell-m)^2,
 \label{eq:Kpm-early}
\end{equation}
their ratio is $\kappa_+/\kappa_-=K_-/K_+$ for the axis labels defined above.  For an azimuthal period $\Delta\phi_A$, the deficits are $\delta_\pm=2\pi-\kappa_\pm\Delta\phi_A$.  Since $K_+-K_-=4m\ell$, both axes can be regularized with one period only if $m\ell=0$.  For $m\ne0$ and $\ell\ne0$, the unequal conicities indicate a string or strut supporting the black hole.  The Weyl rod structure, however, contains no separate acceleration-horizon rod.

Using the freedom to translate the Weyl coordinate $z_A$, we henceforth set $\beta_1=0$ and $\beta_2=\ell$, so that the black-hole rod is centered at $z_A=\ell$.

\subsection{The Ovcharenko--Podolsk\'y solution}
The neutral accelerating Schwarzschild--BR solution of Ovcharenko and Podolsk\'y (OP)~\cite{OP2026} can be written
\begin{equation}
 \dif s^2=\frac1{\Omega^2}\left[-Q(R)\dif t^2+\frac{\dif R^2}{Q(R)}+R^2\left(\frac{\dif x^2}{P(x)(1-x^2)}+P(x)(1-x^2)\dif\phi^2\right)\right],
\label{eq:OPmetric}
\end{equation}
where
\begin{align}
 P(x)&=1-2AMx+B^2M^2x^2,\label{eq:P}\\
 Q(R)&=\left(1-B^2M^2-\frac{2M}{R}\right)\left[1+(B^2-A^2)R^2\right].\label{eq:Q}
\end{align}
Here $M$ is the black-hole scale, $A$ the acceleration parameter, and $B$ the BR field scale.  We denote the OP radial coordinate by $R$, rather than $r$ as in Ref.~\cite{OP2026}.  The conformal factor admits the positive decomposition
\begin{equation}
 \Omega^2=C^2+B^2R^2P(1-x^2),\qquad
 C=1-ARx+B^2MRx^2.
\label{eq:Omega}
\end{equation}
In the magnetic duality frame used here, the same solution is supported by
\begin{equation}
 A^{O}=A_\phi^{O}\,\dif\phi,
 \qquad
 A_\phi^{O}=\frac{1}{B}\left(1+\frac{C}{\Omega}\right)+A_{\phi0}^{O},
\label{eq:OP-potential}
\end{equation}
where $A_{\phi0}^{O}$ is an additive gauge constant; the overall sign depends on the magnetic orientation.

For $M>0$ and $0<BM<1$, the first factor of $Q(R)$ gives the black-hole horizon at
\begin{equation}
 R_H=\frac{2M}{1-B^2M^2}.
\label{eq:RH}
\end{equation}
The second factor, $1+(B^2-A^2)R^2$, has no real zero for $A^2<B^2$.  At $A^2=B^2$ it becomes constant, while for $A^2>B^2$ it has the positive root
\begin{equation}
 R_A=\frac{1}{\sqrt{A^2-B^2}}.
\label{eq:RA}
\end{equation}
The corresponding Weyl horizon structure will be examined below.

\section{Weyl coordinates and rod structure}\label{sec:weyl}
We first examine the horizon rods of the Ovcharenko--Podolsk\'y solution in canonical Weyl coordinates, and then compare its black-hole rod with that of the Alekseev--Astorino solution.

For a static axisymmetric metric, the canonical Weyl radial coordinate is defined by $\rho^2=-\det G$, where $G$ is the Gram matrix of the stationary and axial Killing vectors.  The axial Weyl coordinate $z$ is the corresponding harmonic conjugate on the two-dimensional orbit space.  On a regular static patch, the OP coordinates can be chosen as
\begin{equation}
 \rho_O=\frac{R\sqrt{Q(R)P(x)(1-x^2)}}{\Omega^2},\qquad
 z_O=\frac{[R-M-AMRx][x(1+B^2MR)-AR]}{\Omega^2}.
 \label{eq:OP-weyl}
\end{equation}
Taking the limit $R\to R_H$ from the static region gives $\rho_O\to0$ and
\begin{equation}
 z_O\longrightarrow M\frac{(1+B^2M^2)x-2AM}{1+B^2M^2-2AMx}.
 \label{eq:OP-horizon-latitude}
\end{equation}
On the regular horizon branch, this expression increases from $-M$ to $M$ as $x$ varies from $-1$ to $1$. The black-hole horizon is therefore represented by the finite rod
\begin{equation}
 \mathcal H_b^{O}:\qquad -M\le z_O\le M.
 \label{eq:OP-bh-rod}
\end{equation}

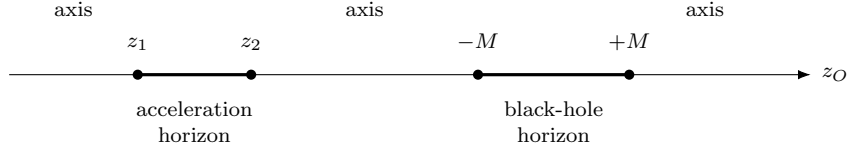
\begin{figure}[H]
\centering
\begin{tikzpicture}[x=1cm,y=1cm,>=Latex,font=\small]
\draw[->,thin] (-5.2,0)--(5.4,0) node[right,font=\scriptsize] {$z_O$};
\draw[very thick] (-3.5,0)--(-2.0,0); \fill (-3.5,0) circle (1.9pt); \fill (-2.0,0) circle (1.9pt);
\node[above=5pt,font=\scriptsize] at (-3.5,0) {$z_1$}; \node[above=5pt,font=\scriptsize] at (-2.0,0) {$z_2$};
\node[below=6pt,font=\scriptsize,align=center] at (-2.75,0) {acceleration\\horizon};
\draw[very thick] (1.0,0)--(3.0,0); \fill (1.0,0) circle (1.9pt); \fill (3.0,0) circle (1.9pt);
\node[above=5pt,font=\scriptsize] at (1.0,0) {$-M$}; \node[above=5pt,font=\scriptsize] at (3.0,0) {$+M$};
\node[below=6pt,font=\scriptsize,align=center] at (2.0,0) {black-hole\\horizon};
\node[above=18pt,font=\scriptsize] at (-4.35,0) {axis}; \node[above=18pt,font=\scriptsize] at (-0.5,0) {axis}; \node[above=18pt,font=\scriptsize] at (4.0,0) {axis};
\end{tikzpicture}
\caption{The separated supercritical OP rod structure.  In addition to the black-hole horizon $[-M,M]$, the Weyl axis contains the finite acceleration-horizon rod $[z_1,z_2]$.  Compare Fig.~\ref{fig:astorino-rod}.}
\label{fig:OP-supercritical-rods}
\end{figure}

The black-hole horizon is present in both representations.  The OP radial function also has an additional real root when $A^2>B^2$, as noted in Sec.~\ref{sec:two}.  To distinguish the corresponding Weyl structure from the single finite horizon rod of the Alekseev--Astorino solution, consider the separated supercritical range
\begin{equation}
 M>0,\qquad 0<BM<1,\qquad
 B<A<\frac{1+B^2M^2}{2M}.
\end{equation}
Writing
\begin{equation}
 s_A=\sqrt{A^2-B^2},
\end{equation}
the positive root lies at $R=1/s_A$.  The limit of \eqref{eq:OP-weyl} on this surface gives $\rho_O=0$ and a second finite rod, whose endpoints are obtained at $x=\pm1$:
\begin{equation}
 \mathcal H_A^+:[z_1,z_2],\qquad
 z_1=-\frac{1}{A-s_A},\qquad
 z_2=-\frac{1}{A+s_A}.
 \label{eq:OP-acc-rod}
\end{equation}
In the stated range, $z_1<z_2<-M$.  The two horizon rods and the intervening axis segment occur in the order
\begin{equation}
 (-\infty,z_1)_{\rm axis}\;|\;[z_1,z_2]_{\mathcal H_A^+}\;|\;(z_2,-M)_{\rm axis}\;|\;[-M,M]_{\mathcal H_b}\;|\;(M,\infty)_{\rm axis}.
 \label{eq:OP-rod-order}
\end{equation}
Here $\mathcal H_A^+$ and $\mathcal H_b$ denote the acceleration and black-hole horizon rods.  The remaining intervals are axis segments.  This ordering is illustrated in Fig.~\ref{fig:OP-supercritical-rods}.

The two canonical Weyl systems admit an arbitrary choice of axial origin and a common constant scale associated with the normalization of the Killing coordinates.  We therefore seek an orientation-preserving affine relation
\begin{equation}
 z_A=c z_O+d,\qquad \rho_A=c\rho_O,\qquad c>0.
\end{equation}
Identifying the two black-hole rods requires the OP endpoints $z_O=\pm M$ and the Astorino endpoints $z_A=\ell\pm m$ to satisfy
\begin{equation}
 -cM+d=\ell-m,\qquad cM+d=\ell+m,
\end{equation}
so that $c=m/M$ and $d=\ell$.  Hence
\begin{equation}
 z_A=\ell+\frac{m}{M}z_O,\qquad
 \rho_A=\frac{m}{M}\rho_O.
\label{eq:weyl-map}
\end{equation}
The affine map identifies the black-hole rods but does not determine the physical parameter correspondence.

\begin{figure}[H]
\centering
\begin{tikzpicture}[x=1cm,y=1cm,>=Latex,font=\small]
\node[font=\scriptsize\sffamily] at (0,1.65) {Astorino after choosing $\beta_1=0$, $\beta_2=\ell$};
\draw[thick] (-4.2,0.85)--(-1.25,0.85); \draw[very thick] (-1.25,0.85)--(1.25,0.85); \draw[thick] (1.25,0.85)--(4.2,0.85);
\node[below,font=\scriptsize] at (-1.25,0.85) {$\ell-m$}; \node[below,font=\scriptsize] at (1.25,0.85) {$\ell+m$};
\draw[->,thick] (0,0.18)--(0,-0.58);
\node[right,font=\scriptsize] at (0.10,-0.20) {$z_A=\ell+(m/M)z_O$, $\rho_A=(m/M)\rho_O$};
\node[font=\scriptsize\sffamily] at (0,-0.95) {OP black-hole rod};
\draw[thick] (-4.2,-1.55)--(-1.25,-1.55); \draw[very thick] (-1.25,-1.55)--(1.25,-1.55); \draw[thick] (1.25,-1.55)--(4.2,-1.55);
\node[below,font=\scriptsize] at (-1.25,-1.55) {$-M$}; \node[below,font=\scriptsize] at (1.25,-1.55) {$+M$};
\end{tikzpicture}
\caption{The first step in testing a relation between the two families: identify the finite black-hole rods that are present in both descriptions.  Their centers fix the Weyl translation and their half-lengths fix the common Weyl scale.  No assumption about the OP acceleration regime is needed at this stage.}
\label{fig:rod-matching}
\end{figure}
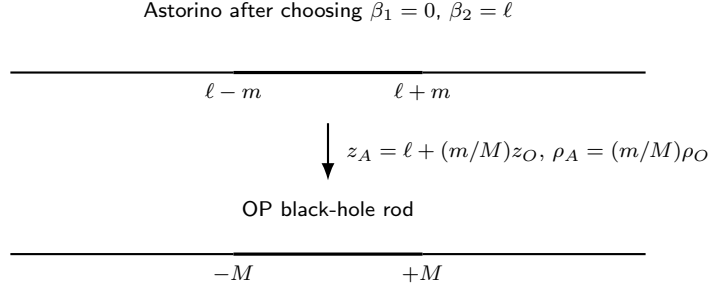

The black-hole rod $[-M,M]$ defines a natural prolate spheroidal system.  The distances from its endpoints are
\begin{equation}
 \mathcal R_\pm^O=\sqrt{\rho_O^2+(z_O\pm M)^2},
\end{equation}
and the corresponding prolate coordinates are defined by
\begin{equation}
 \mathcal R_O=\frac{\mathcal R_+^O+\mathcal R_-^O}{2M},\qquad
 X_O=\frac{\mathcal R_+^O-\mathcal R_-^O}{2M}.
\label{eq:OP-prolate-definition}
\end{equation}
They obey
\begin{equation}
 \rho_O^2=M^2(\mathcal R_O^2-1)(1-X_O^2),\qquad
 z_O=M\mathcal R_O X_O.
\label{eq:OP-prolate-relations}
\end{equation}
Substituting the Weyl transformation \eqref{eq:OP-weyl} and choosing the regular exterior branch gives
\begin{equation}
 \mathcal R_O=\frac{R-M-AMRx}{\Omega},\qquad
 X_O=\frac{x(1+B^2MR)-AR}{\Omega}.
\label{eq:prolate-factor}
\end{equation}
The affine Weyl map \eqref{eq:weyl-map} takes the endpoints $\pm M$ to $\ell\pm m$.  The corresponding Astorino black-hole-centered prolate coordinates are therefore
\begin{equation}
 x_2=\frac{m}{M}\frac{R-M-AMRx}{\Omega},\qquad
 y_2=\frac{x(1+B^2MR)-AR}{\Omega}.
\label{eq:x2y2}
\end{equation}
These satisfy \eqref{eq:rod-spheroid-general} under \eqref{eq:weyl-map} on the same exterior branch.

\section{Horizon geometry and parameter correspondence}\label{sec:horizon}
The Weyl transformation obtained in the preceding section identifies the black-hole horizon rods of the two solutions, but does not establish that their intrinsic geometries agree.  We now compare the horizon two-metrics to determine the correspondence between their parameters.

We express both horizon metrics in coordinates for which the area density is constant,
\begin{equation}
 \dif s_H^2=\mathcal H\left[\frac{\dif y^2}{F(y)}+F(y)\,\dif\phi^2\right].
 \label{eq:equal-area-general}
\end{equation}
Since $\det h=\mathcal H^2$, the area element is $\dif A=\mathcal H\,\dif y\,\dif\phi$; equal intervals of $y$ therefore correspond to equal areas for a fixed azimuthal interval.  We refer to this as the equal-area form.  The function $F(y)$ describes the horizon shape, while $\mathcal H$ sets its area scale.

On the Astorino black-hole horizon, $x_2=m$.  Taking $y=y_2$, the functions entering the induced horizon metric reduce to
\begin{equation}
 S(y)=b^2+(\ell+my)^2,
 \label{eq:S-horizon}
\end{equation}
and
\begin{equation}
 T(y)=b^2+\ell^2+m^2+2m\ell y.
 \label{eq:T-horizon}
\end{equation}
Writing $L=b^2+\ell^2$, these functions become $S(y)=L+2m\ell y+m^2y^2$ and $T(y)=L+m^2+2m\ell y$.  The Astorino horizon metric takes the form \eqref{eq:equal-area-general} with
\begin{equation}
 F_A(y)=\frac{(1-y^2)S(y)}{\sqrt{f_0}\,T(y)^2}.
\label{eq:FA}
\end{equation}
The ratio $S/T^2$ retains the independent combinations $m^2/L$ and $2m\ell/L$.

The Astorino horizon metric is already in equal-area form, with latitude $y=y_2$.  To compare the intrinsic horizon geometries, we transform the OP latitude $x$ to a coordinate $y$ with the same constant-area-density property.  Preserving the pole orientation and requiring the poles to remain at $y=\pm1$, the transformation satisfies
\begin{equation}
 \mathcal H_O\,\frac{\dif y}{\dif x}=\sqrt{\det h^{O}_{ab}},
 \label{eq:equal-area-ode}
\end{equation}
where $h^O_{ab}$ is the OP horizon two-metric in $(x,\phi)$ coordinates and $\mathcal H_O$ is its constant area-density coefficient.  Integration with the pole conditions gives
\begin{equation}
 y=\frac{Xx-Y}{X-Yx},\qquad X=1+B^2M^2,\qquad Y=2AM.
\label{eq:mobius}
\end{equation}
The fractional-linear map is therefore fixed by the equal-area condition, rather than assumed independently.

The corresponding OP horizon shape function is
\begin{equation}
 F_O(y)=(X^2-Y^2)
 \frac{(1-y^2)[1+Yy+(X-1)y^2]}{(X+Yy)^2}.
\label{eq:FO}
\end{equation}
To compare the latitude dependence, divide \eqref{eq:S-horizon} and \eqref{eq:T-horizon} by $L$:
\begin{align}
 \frac{S(y)}{L}&=1+\frac{2m\ell}{L}y+\frac{m^2}{L}y^2,\\
 \frac{T(y)}{L}&=1+\frac{m^2}{L}+\frac{2m\ell}{L}y.
 \label{eq:poly-pattern}
\end{align}
The OP function contains exactly the same pair of polynomial structures,
\begin{equation}
 1+Yy+(X-1)y^2,\qquad X+Yy.
\end{equation}
Matching the nonconstant latitude dependence gives, coefficient by coefficient,
\begin{equation}
 X-1=\frac{m^2}{L},\qquad Y=\frac{2m\ell}{L},
\label{eq:XY-coeff}
\end{equation}
or
\begin{equation}
 X=\frac{L+m^2}{L},\qquad Y=\frac{2m\ell}{L}.
\label{eq:XY}
\end{equation}
Since $X-1=B^2M^2$ and $Y=2AM$, the candidate physical dictionary is
\begin{equation}
 B^2M^2=\frac{m^2}{b^2+\ell^2},\qquad
 AM=\frac{m\ell}{b^2+\ell^2},
\label{eq:dictionary}
\end{equation}
and division of the two scale-free combinations gives
\begin{equation}
 \frac{A}{B}=\frac{\ell}{\sqrt{b^2+\ell^2}}.
\label{eq:AB}
\end{equation}
The horizon shape functions thus determine a candidate parameter correspondence; the remaining normalization is fixed by the off-horizon metric comparison.

The axial conicities provide an independent check of this correspondence.  For the OP solution, the symmetry axes are at $x=\pm1$, where the conicity factors are proportional to $P(\pm1)$.  Thus
\begin{equation}
 P_N=P(+1)=1-2AM+B^2M^2,\qquad
 P_S=P(-1)=1+2AM+B^2M^2.
\end{equation}
Substituting \eqref{eq:dictionary} gives
\begin{equation}
 \frac{P_N}{P_S}=\frac{b^2+(\ell-m)^2}{b^2+(\ell+m)^2}
 =\frac{K_-}{K_+},
\label{eq:conicity}
\end{equation}
in agreement with the Alekseev--Astorino conicity ratio obtained in Sec.~\ref{sec:two}.  The horizon geometry and axial conicities thus give a consistent parameter correspondence between the two representations.

\section{Exact Einstein--Maxwell equivalence}\label{sec:full}
The horizon geometry and axial conicities yield the candidate parameter correspondence, but do not establish equivalence of the four-dimensional solutions.  We now test that correspondence away from the horizon by comparing the complete metric and electromagnetic field.

The affine Weyl map \eqref{eq:weyl-map} already determines the black-hole-centered prolate coordinates $(x_2,y_2)$ through \eqref{eq:x2y2}.  The Alekseev--Astorino solution also depends on the BR-adapted oblate coordinates $(x_1,y_1)$.  To complete the coordinate transformation, we reconstruct these coordinates from the OP solution and verify their defining oblate relations.

The magnetic potentials provide a convenient starting point for this reconstruction.  For the Alekseev--Astorino solution, define
\begin{equation}
 H_A=\frac{2(x_2+my_1)}{(1-y_1)(x_2-m)}.
 \label{eq:HA}
\end{equation}
The lapse and magnetic potential then satisfy
\begin{equation}
 \frac{f_A}{\rho_A^2}=-\frac{H_A^2}{4b^2(H_A-1)},\qquad
 A_{\phi}^{\rm Ast}=\frac{2b}{H_A}+A_{\phi0}^{\rm Ast}.
 \label{eq:A-seed-relations}
\end{equation}
For the OP solution, define
\begin{equation}
 H_O=\frac{2\Omega}{\Omega+C},
 \label{eq:HO}
\end{equation}
so that $A_\phi^O=2/(BH_O)+A_{\phi0}^O$.  The common functional dependence of the potentials suggests $H_A=H_O$ as a candidate reconstruction condition, with the angular normalization to be checked below.  Solving this condition for $y_1$ gives
\begin{equation}
 y_1=-\frac{CD+M\Omega^2}{\Omega(CM+D)},\qquad
 D=R-M-AMRx,
 \label{eq:y1}
\end{equation}
and, on a patch where $y_1\ne0$, the first oblate relation in
\eqref{eq:br-spheroid-general} gives
\begin{equation}
 x_1=\frac{z_A}{y_1}.
 \label{eq:x1}
\end{equation}
The reconstructed coordinates must also satisfy the defining oblate relations (2.2).

To check the oblate relations, we express the Weyl transformation and the reconstructed latitude as rational functions.  Introduce the abbreviations
\begin{equation}
 L=b^2+\ell^2,\qquad s=\frac{R}{M},\qquad
 a=AM=\frac{m\ell}{L},\qquad
 \beta=B^2M^2=\frac{m^2}{L}.
 \label{eq:offhorizon-dimless}
\end{equation}
In terms of these variables, write
\begin{align}
 P&=1-2ax+\beta x^2, &
 C&=1-asx+\beta sx^2,\nonumber\\
 d&=s-1-asx, &
 E&=x(1+\beta s)-as,\nonumber\\
 W&=C^2+\beta s^2P(1-x^2)=\Omega^2.
 \label{eq:offhorizon-rational}
\end{align}
The affine Weyl map and the OP Weyl coordinates then give
\begin{equation}
 \rho_A^2=\frac{m^2s^2QP(1-x^2)}{W^2},\qquad
 z_A=\ell+\frac{mEd}{W},
 \label{eq:offhorizon-weyl}
\end{equation}
where
\begin{equation}
 Q=\left(1-\beta-\frac{2}{s}\right)
 \left[1+(\beta-a^2)s^2\right].
\end{equation}
Equation \eqref{eq:y1} becomes
\begin{equation}
 y_1=-\frac{Cd+W}{\sqrt W(C+d)},\qquad
 Y\equiv y_1^2=\frac{(Cd+W)^2}{W(C+d)^2}.
 \label{eq:offhorizon-Y}
\end{equation}
To test whether the reconstructed pair really defines oblate coordinates, eliminate $x_1$ from $z_A=x_1y_1$ and $\rho_A^2=(b^2+x_1^2)(1-y_1^2)$.  The resulting condition is
\begin{equation}
 \mathcal P_{\rm obl}\equiv
 b^2Y^2+(\rho_A^2+z_A^2-b^2)Y-z_A^2=0.
 \label{eq:Pobl}
\end{equation}
Substitution of \eqref{eq:offhorizon-dimless}--\eqref{eq:offhorizon-Y} into \eqref{eq:Pobl}, followed by reduction to a common denominator, gives an identically vanishing numerator upon imposing the parameter dictionary.  The identity holds wherever the displayed denominators are nonzero.  Together with the unsquared relation \eqref{eq:y1}, this verifies both oblate relations on the selected real branch.
The same condition fails at generic points if $A=0$ is imposed while retaining $\ell\ne0$.  Thus the finite-$\ell$ geometry is not merely a Weyl translation of the nonaccelerating solution.

The quotient in \eqref{eq:x1} is used only where $y_1\ne0$.
The oblate coordinates have Jacobian
\begin{equation}
 \frac{\partial(\rho_A,z_A)}{\partial(x_1,y_1)}
 =\frac{x_1^2+b^2y_1^2}{\rho_A}.
 \label{eq:oblate-jacobian}
\end{equation}
Thus $y_1=0$ does not obstruct local inversion when $x_1\ne0$;
the focal point $x_1=y_1=0$ is excluded from this coordinate patch.
The reconstructed coordinates therefore satisfy
\begin{equation}
 H_A(x_1,y_1,x_2,y_2)=H_O(R,x)
 \label{eq:H-equality}
\end{equation}
throughout the common coordinate patch.

We next compare the orbit-space conformal factors and Killing coefficients.  Substituting the reconstructed coordinates into the Astorino conformal factor, with $K_\pm$ as defined earlier, gives
\begin{equation}
 D_0=\frac{mR}{ML\Omega}K_+K_-.
 \label{eq:D0-factor}
\end{equation}
Writing the OP orbit-space metric in canonical Weyl coordinates as
\begin{equation}
 \dif s_{O,\mathrm{orb}}^2
 =\Lambda_O(\dif\rho_O^2+\dif z_O^2),
\end{equation}
on a common off-horizon static Weyl patch, its coefficient is
obtained from the radial derivatives of the Weyl map:
\begin{equation}
 \Lambda_O=\frac{1}{\Omega^2Q(R)
 [ (\partial_R\rho_O)^2+(\partial_Rz_O)^2 ]}.
 \label{eq:Lambda-OP}
\end{equation}
On the Astorino side, the affine map gives
\begin{equation}
 \Lambda_A=\frac{m^2}{M^2}\frac{b^2f_0D_0^2}
 {\Delta_A(x_2^2-m^2y_2^2)}.
 \label{eq:Lambda-Ast}
\end{equation}
Using \eqref{eq:D0-factor} and the reconstructed spheroidal
coordinates, equality of the orbit-space coefficients $\Lambda_O=\Lambda_A$ determines the remaining constant $f_0$:
\begin{equation}
 f_0=\frac{M^2L^3}{b^2m^2K_+^2K_-^2},
 \label{eq:f0-match}
\end{equation}
and hence
\begin{equation}
 \Lambda_O=\frac{m^2}{M^2}\frac{b^2g_A}{\Delta_A}.
 \label{eq:orbit-conformal-match}
\end{equation}
The equality holds on the common regular off-horizon patch.
Its expression in these coordinates is not used at a zero of $Q(R)$,
where the Weyl formula \eqref{eq:Lambda-OP} requires a limiting
interpretation.

The remaining Killing coefficients agree after constant coordinate rescalings.  In particular,
\begin{equation}
 \frac{f_A}{f_O}=\frac{L}{b^2},
 \label{eq:lapse-ratio}
\end{equation}
and the canonical Weyl determinant fixes the azimuthal coefficient.
With positive orientation choices, take
\begin{equation}
 t_A=\frac{b}{\sqrt L}\,t_O,\qquad
 \phi_A=\frac{M\sqrt L}{mb}\,\phi_O.
 \label{eq:killing-rescalings}
\end{equation}
Finally, the magnetic potentials must agree under the same angular rescaling.  Since $BM=m/\sqrt L$, Eq.~\eqref{eq:H-equality} gives
\begin{equation}
 \frac{2b}{H_A}\,\dif\phi_A
 =\frac{2}{BH_O}\,\dif\phi_O,
 \label{eq:magnetic-potential-match}
\end{equation}
up to additive gauge constants and reversal of the magnetic orientation.
The magnetic Maxwell two-forms therefore agree.

The complete transformation from OP to Alekseev--Astorino
coordinates is therefore as follows.  With $L=b^2+\ell^2$ and the
parameter correspondence
\begin{equation}
 B^2M^2=\frac{m^2}{L},\qquad AM=\frac{m\ell}{L},
 \label{eq:conversion-parameters}
\end{equation}
the two spheroidal systems are obtained directly from $(R,x)$ by
\begin{align}
 x_2&=\frac{m(R-M-AMRx)}{M\Omega}, &
 y_2&=\frac{x(1+B^2MR)-AR}{\Omega},\nonumber\\
 y_1&=-\frac{CD+M\Omega^2}{\Omega(CM+D)}, &
 x_1&=\frac{\ell+x_2y_2}{y_1},
 \label{eq:conversion-coordinates}
\end{align}
where $C=1-ARx+B^2MRx^2$ and $D=R-M-AMRx$; $\Omega$ is the OP
function defined in Sec.~\ref{sec:two}.  The last quotient is used
only where $y_1\ne0$, with the defining oblate relations supplying its
regular continuation where applicable.  The Killing coordinates obey
\begin{equation}
 t_A=\frac{b}{\sqrt L}\,t_O,\qquad
 \phi_A=\frac{M\sqrt L}{mb}\,\phi_O.
 \label{eq:conversion-killing}
\end{equation}
The original electric Alekseev--Astorino field is first transformed to
its magnetic dual, with the orientation and additive gauge constants
chosen as above.

These transformations establish
\begin{equation}
 (g,F)_{\Ast}\cong(g,F)_{\OP}
 \label{eq:EM-equivalence}
\end{equation}
locally on the common regular real patch.  Different choices of
azimuthal period need not give the same global conical quotient.

\section{Physical meaning of the subcritical family}\label{sec:physics}
The exact correspondence identifies the acceleration regime represented by the finite-$\ell$ Alekseev--Astorino solution and gives a physical interpretation of its relative Weyl placement.

\subsection{The subcritical parameter range}
From \eqref{eq:AB},
\begin{equation}
 \frac{A^2}{B^2}=\frac{\ell^2}{b^2+\ell^2}<1
\end{equation}
for every finite real $\ell$ and $b\ne0$.  Conversely, with the positive $b$ and $B$ conventions,
\begin{equation}
 \frac{\ell}{b}=\frac{A}{\sqrt{B^2-A^2}}
 \label{eq:inverse-projective}
\end{equation}
is real throughout $|A|<|B|$.  Thus the finite-real-$\ell$ family covers precisely the subcritical acceleration interval.

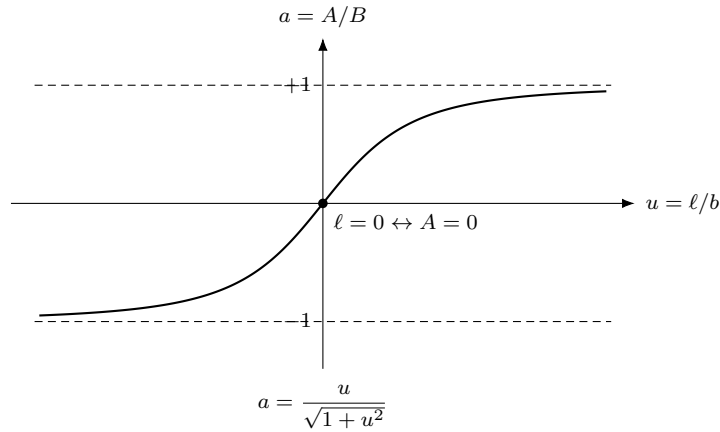
\begin{figure}[H]
\centering
\begin{tikzpicture}[x=1.25cm,y=1.25cm,>=Latex,font=\small]
\draw[->] (-3.3,0)--(3.3,0) node[right,font=\scriptsize] {$u=\ell/b$};
\draw[->] (0,-1.75)--(0,1.75) node[above,font=\scriptsize] {$a=A/B$};
\draw[densely dashed] (-3.05,1.25)--(3.05,1.25); \draw[densely dashed] (-3.05,-1.25)--(3.05,-1.25);
\node[left,font=\scriptsize] at (0,1.25) {$+1$}; \node[left,font=\scriptsize] at (0,-1.25) {$-1$};
\draw[thick,domain=-3:3,samples=100,smooth,variable=\u] plot ({\u},{1.25*\u/sqrt(1+\u*\u)});
\fill (0,0) circle (1.8pt); \node[below right,font=\scriptsize] at (0,0) {$\ell=0\leftrightarrow A=0$};
\node[font=\scriptsize] at (0,-2.15) {$\displaystyle a=\frac{u}{\sqrt{1+u^2}}$};
\end{tikzpicture}
\caption{The map between the finite-real-$\ell$ Alekseev--Astorino family and the subcritical OP acceleration interval.  The critical values $a=\pm1$ are approached only as $|u|\to\infty$.}
\label{fig:parameter-map}
\end{figure}

Writing $u=\ell/b$ and $a=A/B$, we obtain
\begin{equation}
 a(u)=\frac{u}{\sqrt{1+u^2}},\qquad
 \frac{\dif a}{\dif u}=\frac{1}{(1+u^2)^{3/2}}>0.
 \label{eq:monotonic-map}
\end{equation}
The map is a monotonic bijection from $u\in\mathbb R$ to $-1<a<1$, as illustrated in Fig.~\ref{fig:parameter-map}.  The centered member $\ell=0$ corresponds to $A=0$, while the critical values $a=\pm1$ are approached only as $u\to\pm\infty$.  Thus no finite real $\ell$ represents the critical or supercritical regime.

\subsection{Acceleration without an acceleration horizon}
As noted in Sec.~\ref{sec:two}, the second factor of $Q(R)$ has no real root for $A^2<B^2$.  Thus the finite-$\ell$ family has no separate acceleration horizon.  For $\ell\ne0$, however, the parameter correspondence gives $A\ne0$, while the two exterior axes have unequal conicities.  The unequal axial conicities are consistent with a string or strut supporting the accelerating black hole.

This static, conically supported family is distinct from Alekseev's ``geodesic motion'' Schwarzschild--BR construction~\cite{Alekseev2017}.  The result established here is the exact Einstein--Maxwell identification of the finite-$\ell$ solution with the subcritical OP branch, not the general possibility of acceleration without an acceleration horizon.

\subsection{Interpretation in the BR background}
In the pure-background limit $M=0$ of the OP metric, the static worldline $R=0$ is regular and has proper acceleration $|A|$.  For comparison, a static orbit at $\varrho=\varrho_0$ in global coordinates on the $\mathrm{AdS}_2$ factor of BR spacetime has acceleration $B|\varrho_0|/\sqrt{1+\varrho_0^2}$.  Choosing the orientation consistently gives
\begin{equation}
 \varrho_0=\frac{A}{\sqrt{B^2-A^2}}.
\end{equation}
The parameter correspondence therefore implies
\begin{equation}
 \frac{\ell}{b}=\varrho_0.
 \label{eq:rho0}
\end{equation}
Thus $\ell/b$ is the position parameter of the corresponding accelerated orbit in this choice of BR static frame, not a proper distance from a preferred centre.  In particular, $|\varrho_0|$ diverges as $|A/B|\to1^-$.

The dependence of this interpretation on the BR scale can also be seen from the parameter correspondence.  Taking $b\to\infty$ at fixed finite $\ell$ gives
\begin{equation}
 \frac{A}{B}=\frac{\ell}{\sqrt{b^2+\ell^2}}\longrightarrow0.
\end{equation}
This establishes the vanishing of the acceleration-to-background-scale ratio in that limit.  It does not, by itself, determine the separate limits of $A$ and $B$ or prove convergence of the full metric to a vacuum solution.  The physical interpretation of $\ell$ established here is therefore tied to its correspondence with acceleration in the BR background.

\section{Discussion}
We have shown that the finite-real-$\ell$ neutral Alekseev--Astorino solution is locally equivalent to the subcritical accelerating Schwarzschild--Bertotti--Robinson solution of Ovcharenko and Podolsk\'y.  The identification is established by an explicit coordinate and parameter transformation relating the complete metric and electromagnetic field.

The result clarifies the meaning of the relative Weyl placement parameter $\ell$.  Rather than introducing an independent degree of freedom in the neutral family, it parametrizes the subcritical acceleration already present in the Ovcharenko--Podolsk\'y representation.  In particular, finite real $\ell$ covers precisely $|A|<|B|$, where nonzero acceleration is compatible with unequal axial conicities and the absence of a separate acceleration horizon.

The equivalence is local and does not, by itself, identify different global azimuthal periods or their associated conical quotients.  The present analysis is restricted to the neutral sector.

The critical boundary $|A|=|B|$ and the supercritical regime $|A|>|B|$ lie outside the finite-real-$\ell$ parametrization.  Their detailed horizon geometry and the global causal structure of the accelerating Schwarzschild--BR family are beyond the scope of the present work.


\begin{thebibliography}{99}

\bibitem{Bertotti1959} B.~Bertotti, Uniform electromagnetic field in the theory of general relativity, Phys. Rev. \textbf{116}, 1331 (1959).

\bibitem{Robinson1959} I.~Robinson, A solution of the Maxwell--Einstein equations, Bull. Acad. Pol. Sci. Ser. Sci. Math. Astron. Phys. \textbf{7}, 351 (1959).

\bibitem{AlekseevGarcia1996} G.~A.~Alekseev and A.~A.~Garcia, Schwarzschild black hole immersed in a homogeneous electromagnetic field, Phys. Rev. D \textbf{53}, 1853 (1996).

\bibitem{OrtaggioAstorino2018} M.~Ortaggio and M.~Astorino, Ultrarelativistic boost of a black hole in the magnetic universe of Levi--Civita--Bertotti--Robinson, Phys. Rev. D \textbf{97}, 104052 (2018), arXiv:1805.05382.

\bibitem{VanDenBerghCarminati2020} N.~Van den Bergh and J.~Carminati, Non-aligned Einstein--Maxwell Robinson--Trautman fields of Petrov type D, Class. Quantum Grav. \textbf{37}, 215010 (2020).

\bibitem{OP2025} H.~Ovcharenko and J.~Podolsk\'y, New class of rotating charged black holes with nonaligned electromagnetic field, Phys. Rev. D \textbf{112}, 064076 (2025).

\bibitem{OP2026} H.~Ovcharenko and J.~Podolsk\'y, Static black holes in an external uniform electromagnetic field: Reissner--Nordstr\"om black holes accelerating in Bertotti--Robinson spacetime, Phys. Rev. D \textbf{114}, 024068 (2026), arXiv:2602.15462.

\bibitem{OPUniqueness2026} H.~Ovcharenko and J.~Podolsk\'y, Uniqueness of stationary axisymmetric type-D black holes with nonaligned electromagnetic field, Phys. Rev. D \textbf{114}, 044024 (2026), arXiv:2604.13202.

\bibitem{Astorino2025} M.~Astorino, Black holes in the external Bertotti--Robinson--Bonnor--Melvin electromagnetic field, Phys. Rev. D \textbf{112}, 104077 (2025), arXiv:2508.12908.

\bibitem{Astorino2026} M.~Astorino, Reissner--Nordstr\"om in the Bertotti--Robinson--Bonnor--Melvin electromagnetic field, arXiv:2609.00882 (2026).

\bibitem{DiPinto2026} A.~Di Pinto, A more general exact solution for the Schwarzschild black hole in a Bertotti--Robinson--Bonnor--Melvin universe, arXiv:2608.18213 (2026).

\bibitem{Alekseev2017} G.~A.~Alekseev, Solution for ``geodesic'' motion of a Schwarzschild black hole along a magnetic field in $\mathrm{AdS}_2\times S^2$ space-time, arXiv:1512.06289.

\bibitem{Alekseev2025} G.~A.~Alekseev, Charged black hole accelerated by spatially homogeneous electric field of Bertotti--Robinson ($\mathrm{AdS}_2\times S^2$) space-time, arXiv:2511.06082 (2025).

\bibitem{Barrientos2026} J.~Barrientos, A.~Cisterna, A.~D\'iaz, and K.~M\"uller, From Bertotti--Robinson to vacuum: New exact solutions in general relativity via Harrison and inversion symmetries, arXiv:2602.17581 (2026).

\bibitem{DiasLemos2003} O.~J.~C.~Dias and J.~P.~S.~Lemos, Pair of accelerated black holes in an anti-de Sitter background: The AdS C-metric, Phys. Rev. D \textbf{67}, 064001 (2003), arXiv:hep-th/0210065.
\end{thebibliography}
\end{document}